\documentclass[preprintnumbers]{article}
\usepackage[T1]{fontenc}
\usepackage[latin9]{inputenc}
\usepackage{amsmath}
\usepackage{amsthm}
\usepackage{amssymb}
\usepackage{xcolor}

\makeatletter
\numberwithin{equation}{section}
\numberwithin{figure}{section}

\makeatother

\begin{document}
\thispagestyle{empty}
\vspace*{-15mm}

\begin{flushleft}
{\bf OUJ-FTC-5}\\
\end{flushleft}

\vspace{8pt}
\begin{center}
{\Large\bf
	Measurement theory in classical mechanics\\
}
\vspace{7pt}
\baselineskip 36pt
So Katagiri\footnote{So.Katagiri@gmail.com}

\vspace{7pt}

\small{\it Graduate School of Arts and Sciences, The Open University of Japan,
Chiba 261-8586, Japan}

\end{center}

\begin{abstract}

We investigate Measurement theory in classical mechanics in the formulation
of classical mechanics by Koopman and von Neumann (KvN), which uses Hilbert
space. 
We show a difference between classical and quantum mechanics in the "relative interpretation"
 of the state of the target of measurement and the state of the measurement device.
We also derive the uncertainty relation in classical mechanics.

\end{abstract}

\section{Introduction}

In order to discuss the crucial difference between quantum and classical mechanics, it is essential to compare the two with the same formalism.
Bohm described a quantum system in terms of classical mechanics\cite{key-8}, and he found
 that the presence or absence of quantum potential characterizes the difference between a quantum system and a classical system
 \footnote{Quantum potential is defined as follows.
 Rewriting $\psi$ to $\psi=R\exp(iS)$, we obtain a Hamilton-Jacobi-like equation,
 \begin{equation}
 \frac{\partial S}{\partial t}=-\frac{1}{2m}\left(\frac{\partial S}{\partial x}\right)^{2}+V(x)-\frac{\hbar^2}{2m}\frac{\partial^{2}}{\partial x^{2}}\log R(x,t),
 \end{equation}
 where $-\frac{\hbar^2}{2m}\frac{\partial^{2}}{\partial x^{2}}(\log R)$
 is called quantum potential.}.

However, there is a clear difference between a comparison in the quantum system and a comparison in the classical system.
 The difference is the non-commutative nature of the operator.
Unlike classical mechanics, quantum mechanics is described in terms of q-numbers, i.e., non-commutative physical quantities.
 This difference is apparent in measurement theories, such as Heisenberg's uncertainty principle\cite{key-6-1}.

In other words, quantum mechanics is a world of q-numbers. 
On the other hand, classical mechanics is a world of c-numbers, i.e., a commutative world.
 Such an argument, however, is obscured by the fact that the two are in different formalisms.

To clarify this, it is important to rewrite classical mechanics into
quantum mechanics and compare the two.

Koopman and von Neumann\cite{key-29,key-2},
 at the early stages of quantum mechanics, rewrote classical system in the form of quantum mechanics using the KvN equations.
Later, Gozzi and Mauro studied this formalism in detail \cite{key-53,key-54,key-51}.
Here, we will refer to the quantum-theoretical description of classical mechanics by KvN equations as KvN formalism.
Sudarshan discussed the KvN formalism as
a model of quantum-classical interaction\cite{key-55}.

In the KvN formalism,
classical mechanics is described using non-commutative operators,
as in quantum mechanics.
There, classical mechanics is given by
new variables that are non-commutative with respect to position and momentum,
and time evolution is performed by unitary operators consisting of position and momentum and these new variables.

If the new variable introduced is not included in the Hamiltonian, this formalism is equivalent to classical mechanics. Under such an operator formalism, the difference between classical mechanics and quantum mechanics is not in the non-commutative nature of the operators, but in the form of the commutations relations.

To summarize, the normal case can be illustrated by

\begin{align}
\begin{array}{cc}
	\mathrm{QM} & \mathrm{CM}\\
	\mathtt{q\mathchar`-numbers} & \mathtt{c\mathchar`-numbers}
\end{array}
\end{align}
but in Bohm's argument it is

\begin{align}
\begin{array}{cc}
\mathrm{QM(Bohm)} & \mathrm{CM}\\
\mathtt{c\mathchar`-numbers} & \mathtt{c\mathchar`-numbers}
\end{array}
\end{align}
and it is
\begin{align}
\begin{array}{cc}
\mathrm{QM} & \mathrm{CM}(\mathrm{KvN})\\
\mathtt{q\mathchar`-numbers} & \mathtt{q\mathchar`-numbers}
\end{array}
\end{align}
in KvN formalism.

In this paper, we apply quantum measurement theory, which was originally formulated for quantum systems,
 to classical systems and investigate how we describe measurements of classical systems.
 We show a difference between classical and quantum mechanics in the "relative interpretation"
 of the state of the target of measurement and the state of the measurement device.

Next, we derive the uncertainty relation in classical mechanics.
Until now, measurement in classical mechanics
has not been considered enough
\footnote{
Classical measurement theory has been studied in part as classical information theory, 
and the main result is known as Shannon's coding theorem.
This extension to quantum mechanics is still studied\cite{key-Holevo73}\cite{key-Hausladen}\cite{key-Holevo98}\cite{key-Schumacher}\cite{key-Macchiavello},
and reinterpreting Shannon's coding theory in terms of the KvN formalism is  essential  in comparing quantum information theory with classical information theory.
}. Jens, Wilkens, and Lewenstein found
that the formalism of quantum mechanics is useful to other than quantum
mechanics\cite{key-9}. It makes us expect the KvN formalism has
a new meaning and application in classical mechanics.

The structure of this paper is as follows. First, we review the KvN
formalism in Section 2. Next, we extend the KvN formalism to quantum
mechanics and show that this is equivalent to quantum theory. After
reviewing the observation problem using the von Neumann model in Section
4, we discuss in Section 5 the measurement theory in classical mechanics.
In Section 6, we construct the classical mechanics' version of uncertainty
relation. The last chapter will give a summary and discussion.

The appendices include the following. In Appendix A, we comment that
the KvN formalism for free particles can be regarded as a von Neumann
model. Next, in Appendix B, we discuss the von Neumann model in a
formalism that extends the KvN formalism to quantum mechanics, which
is discussed in Section 3. In Appendix C, we introduce the Kraus operator.
In Appendix D, we discuss the case where the initial condition is
the only known probability. In Appendix E, we comment on the Planck operator.
In Appendix F, we describe in detail the evolution of time in the von Neumann model.

\section{The KvN formalism}

This section briefly introduces the KvN formalism. In quantum mechanics,
the commutation relation between the position and momentum operator
of a particle is given by,

\begin{equation}
[\hat{x},\hat{p}]=i\hbar.
\end{equation}

A state can be written $|\psi\rangle$ as an expansion using position
and momentum eigenvalue states,

\begin{equation}
|\psi\rangle=\int dx|x\rangle\langle x|\psi\rangle=\int dp|p\rangle\langle p|\psi\rangle.
\end{equation}

The time evolution of the state is described using the Hamiltonian
operator $H(\hat{x},\hat{p})$ as

\begin{equation}
i\text{\ensuremath{\hbar}}\frac{\partial}{\partial t}|\psi\rangle=\text{\ensuremath{\hat{H}}}|\psi\rangle.
\end{equation}

The wave function (probability amplitude) $\psi(x)$ is a function
of $x$ only, and $\psi(p)$ is its Fourier transform. So it is not
a function on the phase space.

The essence of the KvN formalism is to introduce operators $\hat{\pi}_{x},\hat{\pi}_{p}$
in addition to $\hat{x},\hat{p}$ and require non-commutability between
$\hat{x},\hat{p}$ and $\hat{\pi}_{x},\hat{\pi}_{p}$ while $\hat{x}$ and $\hat{p}$ are made commutative.

\begin{equation}
[\hat{x},\hat{p}]=[\hat{\pi}_{x},\hat{\pi}_{p}]=0,[\hat{x},\hat{\pi}_{x}]=[\hat{p},\hat{\pi}_{p}]=i.
\end{equation}

Since position and momentum are commutative, state $|\psi\rangle$
can be expanded by simultaneous eigenstates of position and momentum
as

\begin{equation}
|\psi\rangle=\int dxdp|x,p\rangle\langle x,p|\psi\rangle.
\end{equation}

That is, in the KvN formalism, the wave function (probability amplitude)
$\psi(x,p)=\langle x,p|\psi\rangle$ is a complex function in phase
space.

It should be noted that $\psi(x,p)$ is not a pseudo-probability like
a Wigner function \cite{key-58} or a Husimi function \cite{key-59},
but a probability amplitude.

By the Fourier transform, $|\psi\rangle$ can be described as

\begin{align}
|\psi\rangle & =\int dxd\pi_{p}|x,\pi_{p}\rangle\langle x,\pi_{p}|\psi\rangle\\
 & =\int d\pi_{x}dp|\pi_{x},p\rangle\langle\pi_{x},p|\psi\rangle=\int d\pi_{x}d\pi_{p}|\pi_{x}.\pi_{p}\rangle\langle\pi_{x},\pi_{p}|\psi\rangle.
\end{align}

In addition, the Liouvillian will be introduced in correspondence
with the Hamiltonian\footnote{Note, the Hamiltonian does not depend on $\hat{\pi}_{x},\hat{\pi}_{p}$.}
\begin{equation}
\hat{L}=\frac{\partial H}{\partial\hat{p}}\hat{\pi}_{x}-\frac{\partial H}{\partial\hat{x}}\hat{\pi}_{p}.
\end{equation}

The KvN equation corresponding to the Schr\"{o}dinger equation is
introduced as 
\begin{equation}
i\frac{\partial}{\partial t}|\psi\rangle=\hat{L}|\psi\rangle.
\end{equation}

By applying $\langle x,p|$ from the left,

\begin{equation}
i\frac{\partial}{\partial t}\psi(x,p,t)=i\frac{\partial H}{\partial p}\frac{\partial\psi(x,p,t)}{\partial x}-i\frac{\partial H}{\partial x}\frac{\partial\psi(x,p,t)}{\partial p}
\end{equation}
is obtained.

This form is the same as the Liouville equation, but in this case,
there is a difference in that $\psi$ is a complex function. In the
KvN formalism, as with quantum mechanics, $\psi(x,p)$ is regarded
as the probability amplitude, and $|\psi(x,p)|^{2}$ is the probability
density in the phase space.

As an example, let us consider a free particle $H(x,p)=\frac{p^{2}}{2m}$
\cite{key-53}. Using the Liouvillian

\begin{equation}
\hat{L}=\frac{\hat{p}}{m}\hat{\pi}_{x},
\end{equation}
the KvN equation yields
\begin{equation}
i\frac{\partial}{\partial t}|\psi\rangle=\frac{\hat{p}}{m}\hat{\pi}_{x}|\psi\rangle.
\end{equation}

Now applying $\langle\pi_{x},p|$ from the left, we obtain

\begin{equation}
i\frac{\partial}{\partial t}\langle\pi_{x},p|\psi\rangle=\frac{p}{m}\pi_{x}\langle\pi_{x},p|\psi\rangle.
\end{equation}

Using proportionality factor $A$, the solution is given by
\begin{equation}
\langle\pi_{x},p|\psi\rangle=Ae^{i\frac{p}{m}\pi_{x}t}.
\end{equation}

By the Fourier transfor for $\pi_{x}$, we obtain

\begin{equation}
\langle x,p|\psi\rangle=\int d\pi_{x}Ae^{i\frac{p}{m}\pi_{x}t-ix\pi_{x}t}=A\delta(x-\frac{p}{m}t).
\end{equation}

Here, if the initial state is $|x_{0},p_{0}\rangle$, it is $A=1$
and

\begin{equation}
\langle x,p|\psi\rangle=\langle x,p|e^{i\hat{L}t}|x_{0},p_{0}\rangle=\delta(x-\frac{p}{m}t).
\end{equation}

This solution reproduces the linear orbit of a free particle in classical
mechanics.

Since KvN is a rewrite of classical mechanics to quantum mechanics
formalism, it is natural that Hamiltonian $H$ only include $x$
and $p$. However, this does not mean that they are two independent
free particles. This can be understood from the form of Louvillian
$\hat{L}$:

\[
\hat{L}=\frac{\partial H}{\partial\hat{p}}\hat{\pi}_{x}-\frac{\partial H}{\partial\hat{x}}\hat{\pi}_{p}.
\]

If $H$ contains a term like $\hat{\pi}_{x}$, then $\hat{L}$ will
contain a term like $\hat{\pi}_{x}^{2}$. This represents fluctuations,
as is often the case with operator forms in thermodynamics\cite{key-gravityAnalogy}.

In quantum mechanics, this fluctuation is equivalent to adding a quantum
effect. Our argument is to discuss classical mechanics in the form
of quantum mechanics, and does not include such a term.

However, expressing the quantum effect by adding $\pi_{x},\pi_{p}$
are interesting as a way of discussing the boundary region between
classical mechanics and quantum mechanics, and is being studied as
generalized classical mechanics\cite{key-51}.

\section{Relation to quantum mechanics}

Now we consider the relationship between the KvN formalism and quantum
mechanics.

In quantum mechanics, position $\hat{x}_{q}$ and momentum $\hat{p}_{q}$
satisfy the canonical commutation relation,

\begin{equation}
[\hat{x}_{q},\hat{p}_{q}]=i\hbar.
\end{equation}

The same algebra can be constructed using $\hat{x},\hat{\pi}_{x},\hat{p},\hat{\pi}_{p}$.
If we define operators $\hat{x}_{\hbar}$ and $\hat{p}_{\hbar}$ as
\begin{equation}
\hat{x}_{\hbar}=\hat{x}-\frac{1}{2}\hbar\hat{\pi}_{p},
\end{equation}

\begin{equation}
\hat{p}_{\hbar}=\hat{p}+\frac{1}{2}\hbar\hat{\pi}_{x},
\end{equation}
then we obtain

\begin{equation}
[\hat{x}_{\hbar},\hat{p}_{\hbar}]=i\hbar.
\end{equation}

A Similar algebra is discussed in \cite{key-1}\footnote{These equations satisfy the Weyl relation,

\begin{equation}
\hat{U}(t)\hat{V}(s)=e^{ist/\hbar}\hat{V}(s)\hat{U}(t),\hat{U}(t)\hat{U}(t')=\hat{U}(t')\hat{U}(t),\hat{V}(s)\hat{V}(s')=\hat{V}(s')\hat{V}(s),
\end{equation}
\begin{equation}
\hat{U}(t)=e^{i\hat{x}_{h}t/\hbar}=e^{(\hat{x}+\hbar\frac{1}{2}\hat{\pi}_{p})t/\hbar},\hat{V}(s)=e^{i\hat{p}_{\hbar}s/\hbar}=e^{(\hat{p}-\hbar\frac{1}{2}\hat{\pi}_{x})s/\hbar}.
\end{equation}

Then, from the Stone-von Neumann theorem, There exists a unitary
transformation $\hat{U}$ such that
\begin{equation}
\hat{U}e^{i\hat{x}_{\hbar}t/\hbar}=e^{i\hat{x}_{q}t/\hbar}\hat{U},
\end{equation}

\begin{equation}
\hat{U}e^{i\hat{p}_{\hbar}t/\hbar}=e^{i\hat{p}_{q}t/\hbar}\hat{U}.
\end{equation}

Then, $\hat{x}_{\hbar},\hat{p}_{\hbar}$ describe quantum mechanics
exactly.}
\footnote{
We also obtain
\begin{equation}
\hat{\pi}_{x,\hbar}=\hat{x}+\frac{1}{2}\hbar\hat{\pi}_{p},
\end{equation}

\begin{equation}
\hat{\pi}_{p,\hbar}=\hat{p}-\frac{1}{2}\hbar\hat{\pi}_{x}
\end{equation}

and should discuss for the back reaction from these operators.
However, these $\hat{\pi}_{x,\hbar},\hat{\pi}_{p,\hbar}$ are commutative
with $\hat{x}_{\hbar},\hat{p}_{\hbar}$ .
\begin{equation}
[\hat{x}_{\hbar},\hat{\pi}_{x,\hbar}]=0,
\end{equation}

\begin{equation}
[\hat{x}_{\hbar},\hat{\pi}_{p,\hbar}]=0,
\end{equation}

\begin{equation}
[\hat{p}_{\hbar},\hat{\pi}_{x,\hbar}]=0,
\end{equation}

\begin{equation}
[\hat{p}_{\hbar},\hat{\pi}_{p,\hbar}]=0.
\end{equation}

Therefore, it is understood that the back reaction from $\hat{\pi}_{x,\hbar},\hat{\pi}_{p,\hbar}$
does not need to be considered.
}.

Then we obtain quantum wave functions in phase space,

\begin{equation}
i\hbar\frac{\partial}{\partial t}|\psi\rangle=H(\hat{x}+\hbar\frac{1}{2}\hat{\pi}_{p},\hat{p}+\hbar\frac{1}{2}\hat{\pi}_{x})|\psi\rangle.\label{eq:quantumKvn}
\end{equation}

To recover the original KvN formalism, we expand the right hand side
of (\ref{eq:quantumKvn}) in power of $\hbar$, 
\begin{align}
i\frac{\partial}{\partial t}|\psi\rangle & =\frac{1}{\hbar}H|\psi\rangle+\frac{1}{2}\left(-\frac{\partial H(\hat{x},\hat{p})}{\partial\hat{x}}\hat{\pi}_{p}+\frac{\partial H(\hat{x},\hat{p})}{\partial\hat{p}}\hat{\pi}_{p}\right)\\
 & -\frac{\hbar}{2\cdot2^{2}}\left(\frac{\partial^{2}H(\hat{x},\hat{p})}{\partial\hat{x}^{2}}\hat{\pi}_{p}^{2}+\frac{\partial^{2}H(\hat{x},\hat{p})}{\partial\hat{p}^{2}}\hat{\pi}_{x}^{2}+2\frac{\partial^{2}H(\hat{x},\hat{p})}{\partial\hat{x}\partial\hat{p}}\hat{\pi}_{x}\hat{\pi}_{p}\right)+\text{\ensuremath{\dotsb}}.
\end{align}

Then, in the limit $\hbar\to0$ this equation returns to the KvN formalism.

We comment state $|x_{q}\rangle$ in quantum mechanics corrensponds
to $|x,\pi_{p}\rangle$, not $|x,p\rangle$. Therefore, $|x,\pi_{p}\rangle$
and $|p,\pi_{x}\rangle$ in the KvN formalism have a connection with
quantum theory in spite of classical mechanics.

\section{The von Neumann model as a Measurement theory of quantum mechanics}

\subsection{The von Neumann model}

In this section, we introduce the von Neumann model as a simple example
of the measurement model\cite{key-11}.

The system consists of a measurement target and a measurement device,
and the corresponding physical quantities $\{\hat{x},\hat{p}\},\{\hat{X},\hat{P}\}$
satisfy the canonical commutation relations,

\begin{equation}
[\hat{x},\hat{p}]=i\hbar,
\end{equation}

\begin{equation}
[\hat{X},\hat{P}]=i\hbar,
\end{equation}
and the other commutators of $\hat{x},\hat{p},\hat{X},\hat{P}$ are
$0$.

As an interaction between the measurement target and the measurement
device, we introduce a Hamiltonian
\begin{equation}
\hat{H}=\hat{x}\hat{P}
\end{equation}
and the free Hamiltonian part is not considered for the sake of simplicity.
Also, we take $t=1$.

Then, the time evolution operator is given by

\begin{equation}
\hat{U}=e^{-i\hat{x}\hat{P}}.
\end{equation}

Taking the initial state as 

\begin{equation}
|\psi\rangle=|\phi\rangle\otimes|\eta\rangle,
\end{equation}
where $|\phi\rangle$ is the initial state of the measurement target
and $|\eta\rangle$ is the initial state of the measurement device.

The time evolution of state is expressed as

\begin{equation}
|\psi_{\mathrm{after}}\rangle\equiv\hat{U}|\psi\rangle=\int dx|x\rangle\langle x|\phi\rangle\otimes e^{-ix\hat{P}}|\eta\rangle.\label{eq:neum}
\end{equation}

We expand $|\eta\rangle$ using $|X\rangle$ to obtain

\begin{equation}
\begin{aligned}|\psi_{\mathrm{after}}\rangle & =\int dxdX\langle x|\phi\rangle\langle X-x|\eta\rangle|x\rangle\otimes|X\rangle.\end{aligned}
\end{equation}

Next, we perform a projective measurement on the measurement device.
The probability that the measurement device (or needle) obtains $x_{0}$
is
\begin{equation}
\int dx\left|\langle x|\phi\rangle\langle x-x_{0}|\eta\rangle\right|^{2}.
\end{equation}

If the initial state of the measurement device is $|0\rangle_{X}$,
this probability is
\begin{equation}
\int dx\left|\langle x|\phi\rangle\langle x-x_{0}|0\rangle_{X}\right|^{2}=\left|\phi(x_{0})\right|^{2}.
\end{equation}

This equation is consistent with the results of the projective measurement
of the measurement target\cite{key-19}.

\subsection{Relative state}

Because the projection hypothesis is an inherent problem in quantum mechanics, it is conceptually difficult to consider a measurement theory that includes the projection hypothesis in classical mechanics. 
Therefore, in this section, we will introduce the relative state\footnote{This state is discussed by Everett \cite{key-3}.}.

We use the notation
\begin{equation}
|\eta[x]\rangle\equiv e^{-ix\hat{P}}|\eta\rangle.
\end{equation}
Then (\ref{eq:neum}) is rewriten as

\begin{equation}
|\psi_{\mathrm{after}}\rangle=\int dx\langle x|\phi\rangle|x\rangle\otimes|\eta[x]\rangle.
\end{equation}

$|x\rangle\otimes|\eta[x]\rangle$ is called relative state. In relative
state interpretation, $|x\rangle\otimes|\eta[x]\rangle$ is interpreted
as the measurement device observing its position as $x$ \footnote{In our discussion, we do not consider decoherence \cite{key-57} because
we do not adopt the many-worlds interpretation.}.

On the other hand, $|\psi_{\mathrm{after}}\rangle$ can be expanded
as follows,

\begin{equation}
|\psi_{\mathrm{after}}\rangle=\int dP\langle P|\eta\rangle|\phi[P]\rangle\otimes|P\rangle.
\end{equation}

This is different from the previous one, and it can be interpreted
that the measurement target observed the momentum of the measurement
device as $P$.

Note that these two propositions do not hold in relative state at
the same time.

In contrast, these two propositions will hold in relative state at
the same time in measurement theory in classical mechanics.

\section{Measurement theory in classical mechanics}

In this section, we discuss measurement theory in classical mechanics
using the KvN formalism and the von Neumann model.

As in the previous section, the system consists of a measurement target
and a measurement device, and the corresponding physical quantities
$\{\hat{x},\hat{p},\hat{\pi}_{x},\hat{\pi}_{p}\}$, $\{\hat{X},\hat{P},\hat{\pi}_{X},\hat{\pi}_{P}\}$
satisfy the canonical commutation relations,

\begin{equation}
[\hat{x},\hat{p}]=0,[\hat{x},\hat{\pi}_{x}]=i,[\hat{p},\hat{\pi}_{p}]=i,
\end{equation}

\begin{equation}
[\hat{X},\hat{P}]=0,[\hat{X},\hat{\pi}_{X}]=i,[\hat{P},\hat{\pi}_{P}]=i.
\end{equation}
Note that the dimension of $\hat{\pi}_{x}$ and $\hat{\pi}_{p}$ is
$[x^{-1}]$ and $[p^{-1}]$.

From the von Neumann model's Hamiltonian $\hat{H}=\hat{x}\hat{P}$,
we obtain the Liouvillian

\begin{align}
\hat{L} & =\frac{\partial\hat{H}}{\partial\hat{p}}\hat{\pi}_{x}-\frac{\partial\hat{H}}{\partial\hat{x}}\hat{\pi}_{p}+\frac{\partial\hat{H}}{\partial\hat{P}}\hat{\pi}_{X}-\frac{\partial\hat{H}}{\partial\hat{X}}\hat{\pi}_{P}\\
 & =\hat{P}\hat{\pi}_{p}-\hat{x}\hat{\pi}_{X}
\end{align}

Then, the time evolution operator is given by
\footnote{
	The time evolution is set to $t=1$, but the situation is the same for
	Neumann's measurement model in quantum theory. See Appendix F for details of this model.
}

\begin{equation}
\hat{U}=e^{-i\left(\hat{P}\hat{\pi}_{p}-\hat{x}\hat{\pi}_{X}\right)}.
\end{equation}

As with quantum mechanics, the time evolution of state is obtained
as

\begin{align}
|\psi_{\mathrm{after}}\rangle & =e^{-i\left(\hat{P}\hat{\pi}_{p}-\hat{x}\hat{\pi}_{X}\right)}|\phi\rangle\otimes|\eta\rangle\\
 & =\int dx\:dP|\phi(p\to p-P)\rangle\otimes|\eta(X\to X+x)\rangle\\
 & =\int dp\:dx\:dX\:dP\phi(x,p)\eta(X,P)|x,p[-P]\rangle\otimes|X[x],P\rangle
\end{align}
where 
\begin{equation}
|\phi(p\to p-P)\rangle\equiv\int dp\phi(x,p)|x,p[-P]\rangle,
\end{equation}

\begin{equation}
|\eta(X\to X+x)\rangle\equiv\int dX\eta(X,P)|X[x],P\rangle.
\end{equation}

The important difference from quantum mechanics is that in relative
state interpretation, two propositions
\begin{enumerate}
\item the measurement device observed the position of the measurement target
as $x$,
\item the measurement target observed the momentum of the measurement device
as $P$,
\end{enumerate}
hold in relative state at the same time.\footnote{As a more modern approach, we discuss the Kraus operator in classical
mechanics is discussed in Appendix C. It has not been discussed in
KvN formalism until now.}
\footnote{
In classical mechanics, the Hamiltonian $H$ of the von Neumann model
is
\begin{equation}
H=xP,
\end{equation}

which definitely has a non-trivial interaction between $x$ and $P$.
Therefore, the measurement target and the measurement device have
a certain interaction with each other. If it uses quantum mechanical
formalism, the Liouvillian $\hat{L}$ of the von Neumann model is

\begin{equation}
\hat{L}=\hat{P}\hat{\pi}_{p}-\hat{x}\hat{\pi}_{X},
\end{equation}

as if there is no interaction between $\hat{x}$ and $\hat{P}$, and
as a result, simultaneous observation is possible.
This is a non-trivial result specific to the KvN formalism.
}

\section{Uncertainty relations in classical mechanics}

\subsection{Uncertainty relation in classical mechanics with $\hat{x}$ and $\hat{p}$}

Here, we investigate the relationship between error and disturbance.
The Ozawa's inequality is a relational expression for error and disturbance
\cite{key-10}. We discuss how we can obtain the Ozawa's inequality
in classical mechanics.

We introduce error operator $\hat{N}(t)=\hat{X}(t)-\text{\ensuremath{\hat{x}}}$
and disturbance operator $\hat{D}(t)=\hat{p}(t)-\hat{p}$.

The error $\epsilon$ and disturbance $\eta$ are defined by
\begin{equation}
\epsilon=\sqrt{\langle\hat{N}^{2}\rangle}\geq\sigma(\hat{N}),
\end{equation}

\begin{equation}
\eta=\sqrt{\langle\hat{D}^{2}\rangle}\geq\sigma(\hat{D}),
\end{equation}
where $\sigma^{2}(\hat{A})\equiv\langle\hat{A}^{2}\rangle-\langle\hat{A}\rangle^{2}$.
Using Kennard-Robertson uncertainty relations \cite{key-7-1,key-8-1}

\begin{equation}
\sigma(\hat{N})\sigma(\hat{D})\geq\frac{1}{2}\langle[\hat{N},\hat{D}]\rangle,
\end{equation}
we get
\begin{equation}
\epsilon\eta\geq\frac{1}{2}\langle[\hat{N},\hat{D}]\rangle.
\end{equation}

Next, $[\hat{N}(t),\hat{D}(t)]$ is calculated as
\begin{equation}
[\hat{N}(t),\hat{D}(t)]=[\hat{X}(t),\hat{p}(t)]-[\hat{X}(t),\hat{p}]-[\hat{x},\hat{p}(t)]+[\hat{x},\hat{p}].
\end{equation}

Under the reasonable assumption $[\hat{X}(t),\hat{p}(t)]=0$, this
equation yields

\begin{equation}
\langle[\hat{N},\hat{D}]\rangle+\langle[\hat{X}(t),\hat{p}]+[\hat{x},\hat{p}(t)]\rangle=\langle[\hat{x},\hat{p}]\rangle.
\end{equation}

Since $\hat{x}$ and $\hat{p}$ commute in classical mechanics, we
get

\begin{equation}
\langle[\hat{N},\hat{D}]\rangle+\langle[\hat{X}(t),\hat{p}]\rangle+\langle[\hat{x},\hat{p}(t)]\rangle=0
\end{equation}
and its time differentiation
\begin{equation}
\frac{d}{dt}\langle[\hat{N},\hat{D}]\rangle+\langle[\dot{X}(t),\hat{p}]\rangle+\langle[\hat{x},\dot{\hat{p}}(t)]\rangle=0.
\end{equation}

Because the Liouvillian is given by

\begin{align}
\hat{L} & =\frac{\partial\hat{H}}{\partial\hat{p}}\hat{\pi}_{x}-\frac{\partial\hat{H}}{\partial\hat{x}}\hat{\pi}_{p}+\frac{\partial\hat{H}}{\partial\hat{P}}\hat{\pi}_{X}-\frac{\partial\hat{H}}{\partial\hat{X}}\hat{\pi}_{P},
\end{align}
we get 
\begin{equation}
\langle[\dot{\hat{X}}(t),\hat{p}]\rangle+\langle[\hat{x},\dot{\hat{p}}(t)]\rangle=0.
\end{equation}

After all, we get
\begin{equation}
\langle[\hat{N},\hat{p}]\rangle+\langle[\hat{x},\hat{D}]\rangle=-C,
\end{equation}
\begin{equation}
\langle[\hat{N},\hat{D}]\rangle=C
\end{equation}
where $C$ is some time-invariant constant. Then, for $t=0$, 
\begin{equation}
\langle[\hat{N}(0),\hat{D}(0)]\rangle=0=C.
\end{equation}

Then we obtain
\begin{equation}
\langle[\hat{N},\hat{p}]\rangle=\langle[\hat{D},\hat{x}]\rangle,\langle[\hat{N},\hat{D}]\rangle=0.\label{eq:classical}
\end{equation}

Note that

\begin{equation}
\epsilon\eta\geq0,\epsilon\sigma(\hat{p})+\eta\sigma(\hat{x})\geq0\label{eq:cl_ozawa}
\end{equation}
are trivial inequalities. Now, we discuss the condition of the equal
sign.

From the Heisenberg equation for $\hat{L}=\hat{P}\hat{\pi}_{p}-\hat{x}\hat{\pi}_{X}$,
we obtain
\begin{equation}
\frac{d\hat{x}}{dt}=0,\frac{d\hat{p}}{dt}=-\hat{P},
\end{equation}

\begin{equation}
\frac{d\hat{X}}{dt}=\hat{x},\frac{d\hat{P}}{dt}=0,
\end{equation}

\begin{equation}
\frac{d\hat{\pi}_{x}}{dt}=\hat{\pi}_{X},\frac{d\hat{\pi}_{p}}{dt}=0,
\end{equation}

\begin{equation}
\frac{d\hat{\pi}_{X}}{dt}=0,\frac{d\hat{\pi}_{P}}{dt}=-\hat{\pi}_{p}.
\end{equation}

Therefore, with the initial condition of $\hat{x}(0)=\hat{x}_{0},\,\hat{p}(0)=\hat{p}_{0},\,\hat{\pi}_{x}(0)=\hat{\pi}_{x0},\,\hat{\pi}_{p}(0)=\hat{\pi}_{p0}$,
$\hat{X}(0)=\hat{X}_{0},\,\hat{P}(0)=\hat{P}_{0},\,\hat{\pi}_{X}(0)=\hat{\pi}_{X0},\,\hat{\pi}_{P}(0)=\hat{\pi}_{P0}$,
we obtain

\begin{equation}
\hat{x}(t)=\hat{x}_{0},\hat{p}(t)=\hat{p}_{0}-t\hat{P}_{0},
\end{equation}

\begin{equation}
\hat{X}(t)=\hat{X}_{0}+t\hat{x}_{0},\hat{P}(t)=\hat{P}_{0},
\end{equation}

\begin{equation}
\hat{\pi}_{x}(t)=\hat{\pi}_{x0}+t\hat{\pi}_{X0},\hat{\pi}_{p}(t)=\hat{\pi}_{p0},
\end{equation}

\begin{equation}
\hat{\pi}_{X}(t)=\hat{\pi}_{X0},\hat{\pi}_{P}(t)=\hat{\pi}_{P0}-t\hat{\pi}_{p0}.
\end{equation}

Using these results, $\hat{N}$ and $\hat{D}$ are expressed as
\begin{equation}
\hat{N}(t)=\hat{X}_{0}+(t-1)\hat{x}_{0},
\end{equation}

\begin{equation}
\hat{D}(t)=-t\hat{P}_{0}.
\end{equation}

These equations are the same as the result obtained by the von Neumann
model of quantum mechanics.

Then,
\begin{equation}
\langle x,p,X,P|\hat{N}(t)|x,p,X,P\rangle=X+(t-1)x,
\end{equation}

\begin{equation}
\langle x,p,X,P|\hat{D}(t)|x,p,X,P\rangle=-tP.
\end{equation}

Therefore the unbiased condition is given by
\begin{equation}
X=(1-t)x, P=0.
\end{equation}

Under this condition,

\begin{equation}
\langle x,p,X,P|\hat{N}(t)^{2}|x,p,X,P\rangle=(X+(t-1)x)^{2}=0,
\end{equation}

\begin{equation}
\langle x,p,X,P|\hat{D}(t)^{2}|x,p,X,P\rangle=t^{2}P^{2}=0.
\end{equation}

Then, in this condition, we get
\begin{equation}
\epsilon=\eta=0,
\end{equation}

\begin{equation}
\sigma(p)=\sigma(x)=0.
\end{equation}

If the unbiased condition is not satisfied, we get at $t=1,$
\begin{equation}
\epsilon=X,\eta=P.
\end{equation}

It represents the initial calibration of the device. In such a case,
$X=0$ or $P=0$ is the condition of the equal sign of (\ref{eq:cl_ozawa}).

The case where the initial condition is the only known probability
is discussed in Appendix D.

\subsection{Uncertainty relationsin classical mechanics with $\hat{\pi}_{x}$
and $\hat{\pi}_{p}$}

We clarify the role of $\hat{\pi}_{x}$ and $\hat{\pi}_{p}$ in classical
mechanics. Although $\hat{\pi}_{x}$ and $\hat{\pi}_{p}$ are hidden
variables in classical mechanics, they can be expressed as physical
quantities by combining quantum mechanics and classical mechanics.

As confirmed in Section 3, $\hat{\pi}_{x}$ and $\hat{\pi}_{p}$ are
described by

\begin{equation}
\hat{\pi}_{x}=-\frac{2}{\hbar}(\hat{p}_{\hbar}-\hat{p}),
\end{equation}

\begin{equation}
\hat{\pi}_{p}=\frac{2}{\hbar}(\hat{x}_{\hbar}-\hat{x}).
\end{equation}

In these relational expressions, we can determine $\hat{\pi}_{x}$
and $\hat{\pi}_{p}$ by using both classical and quantum observables
\footnote{
Note that these equations can also be regarded as differentiation
by Planck's constant.
This helps us understand the relationship between KvN formalism and quantum theory by treating the derivative of Planck's constant as an operator.
See Appendix E.
}.

Therefore, in addition to the usual disturbance $\hat{D}$, we should
consider another disturbance,
\begin{equation}
\hat{D}_{\pi_{x}}(t)=\hat{\pi}_{x}(t)-\hat{\pi}_{x}
\end{equation}

By a similar argument such as

\begin{equation}
[\hat{N}(t),\hat{D}_{\pi_{x}}(t)]=[\hat{X}(t),\hat{\pi}_{x}(t)]-[\hat{X}(t),\hat{\pi}_{x}]-[\hat{x},\hat{\pi}_{x}(t)]+[\hat{x},\hat{\pi}_{x}],
\end{equation}
using Kennard-Robertson uncertainty relations, non-commutativity of
$\hat{x}$ and $\hat{\pi}_{x}$ gives an Ozawa-like inequality

\begin{equation}
\epsilon\eta_{\pi_{x}}+\epsilon\sigma(\hat{\pi}_{x})+\ensuremath{\sigma}(\ensuremath{\hat{x}})\text{\ensuremath{\eta}}_{\pi_{x}}\geq\frac{1}{2},
\end{equation}
where $\eta_{\pi_{x}}=\sqrt{\langle\hat{D}_{\pi_{x}}\rangle}$. Since
the Planck constant does not appear in this inequality, it holds even
in the classical mechanical limit.

\section{Discussion}

We constructed the measurement theory of classical mechanics.

In contrast to quantum mechanics, we have found two propositions hold
in relative state at the same time.
\begin{enumerate}
\item The measurement device observed the position of the measurement target
as $x$.
\item The measurement target observed the momentum of the measurement device
as $P$.
\end{enumerate}
This difference in simultaneity corresponds to the result of the discussion
of uncertainty relations, and Ozawa's inequality becomes trivial in
classical mechanics. If the initial state is not well known, we can
obtain a relational expression about the error and the disturbance
in the von Neumann model.

We extended the KvN formalism to quantum theory and determined $\hat{\pi}_{x}$
and $\hat{\pi}_{p}$ using both classical and quantum observables.
Then, we also introduced another disturbance on $\hat{\pi}_{x}$ and obtained an Ozawa-like uncertainty relation. Since this relation
is independent of Planck's constant, it holds in classical mechanics.
Treating $\pi_x$ and $\pi_p$ as observables is known as generalized classical mechanics\cite{key-51}.
In this case, the Louvillian becomes an observable, and its eigenvalues are closely related the conditions in ergodic theory\cite{key-Arnol'd}.
As already mentioned, $\pi_x$ and $\pi_p$ are closely related to the effect of quantum fluctuation.
This uncertainty relation may be significant in the theory of intermediate scale
between classical theory and quantum theory.

The application of these relational expressions to behavioral economics
in recent years is astonishing. Through these applications, the role
of phase in classical mechanics may be newly understood.

As an application of measurement theory in classical mechanics, it
is possible to analytically formulate thought experiments in classical
mechanics such as Maxwell's demon and Einstein's optical clock\cite{key-5-1}.
Although many have discussed these in the past, our study can contribute
to the conceptual discussion of science. Further research will reveal
them.

\section*{Appendix A. Measurement interpretation of classical mechanics}

We comment that the KvN formalism for a free particle can be regarded
as a von Neumann model.

In such a case, the Liouvillian is give by
\begin{equation}
\hat{L}=-\frac{\hat{p}}{m}\hat{\pi}_{x}.
\end{equation}

If we regard $|p,x\rangle$ as a composite of the measurement target
$|p\rangle$ and the measurement device $|x\rangle$, the time evolution
of the free particle
\begin{equation}
e^{-i\frac{\hat{p}}{m}\hat{\pi}_{x}t}|p\rangle|x\rangle=|p\rangle|x+\frac{p}{m}t\rangle
\end{equation}
can be regarded as the obseravation of position $x$ by momentum $p$.

Note that the disturbance in this case is $\hat{D}_{\hat{\pi}_{p}}=\hat{\pi}_{p}(t)-\hat{\pi}_{p}$.
$\hat{D}_{\hat{\pi}_{p}}$ is defined in Section 6.

\section*{Appendix B. Measurement theory in the extended KvN formalism}

We discuss the von Neumann model in a formalism that extends the KvN
formalism to quantum mechanics, which is discussed in Section 3.

The Hamiltonian of the von Neumann model is

\begin{equation}
\hat{H}=\hat{x}_{\hbar}\hat{P}_{\hbar}=(\hat{x}+\frac{\hbar}{2}\hat{\pi}_{p})(\hat{P}-\frac{\hbar}{2}\hat{\pi}_{X})=\hat{x}\hat{P}+\frac{\hbar}{2}(\hat{\pi}_{p}\hat{P}-\hat{x}\hat{\pi}_{X})-\frac{\hbar^{2}}{4}\hat{\pi}_{p}\hat{\pi}_{X}.
\end{equation}

We take $|x,\pi_{p}\rangle\otimes|X,P\rangle$ as the initial condition.
We assume in this initial condition that the measurement target is
quantum, and the measurement device is classical.

Time evolution of state is
\begin{align}
e^{i\hat{H}t/\hbar}|x,\pi_{p}\rangle\otimes|X,P\rangle & =e^{ixPt/\hbar+i\frac{1}{2}(\pi_{p}P-x\hat{\pi}_{X})t-\frac{\hbar}{4}\pi_{p}\hat{\pi}_{X}t}|x,\pi_{p}\rangle\text{\ensuremath{\otimes}}|X,P\rangle\\
 & =e^{i(x+i\frac{1}{2}\pi_{p})Pt}|x,\pi_{p}\rangle\otimes|X+\frac{1}{2}(x+\frac{\hbar}{2}\pi_{p})t,P\rangle\\
 & =e^{ix_{\hbar}Pt}|x,\pi_{p}\rangle\otimes|X+\frac{1}{2}x_{\hbar}t,P\rangle.
\end{align}

Next, we take $|x,\pi_{p}\rangle\otimes|\pi_{X},P\rangle$ as the
initial condition. We assume in this initial condition that the measurement
target is quantum, and the measurement device is also quantum.

Time evolution of state is
\begin{align}
e^{i\hat{H}t/\hbar}|x,\pi_{p}\rangle\otimes|X,\pi_{P}\rangle & =e^{ix\hat{P}t/\hbar+i\frac{1}{2}(\pi_{p}\hat{P}-x\hat{\pi}_{X})t-\frac{\hbar}{4}\pi_{p}\hat{\pi}_{X}t}|x,\pi_{p}\rangle\otimes|X,\pi_{P}\rangle\\
 & =e^{i(x+i\frac{1}{2}\pi_{p})Pt}|x,\pi_{p}\rangle\otimes|X+\frac{1}{2}(x+\frac{\hbar}{2}\pi_{p})t|\pi_{P}+(x+\frac{1}{2}\hbar\pi_{p})/\hbar\rangle\\
 & =e^{ix_{\hbar}Pt}|x,\pi_{p}\rangle\otimes|X+\frac{1}{2}x_{\hbar}t,\pi_{P}+x_{\hbar}/\hbar\rangle.
\end{align}

Note that $|X+\frac{1}{2}x_{\hbar}t,\pi_{P}+x_{\hbar}/\hbar\rangle$
is an eigenstate of $\hat{X}_{\hbar}$,
\begin{align}
\hat{X}_{\hbar}|X+\frac{1}{2}x_{\hbar}t,\pi_{P}+x_{\hbar}/\hbar\rangle & =(\hat{X}+\frac{\hbar}{2}\hat{\pi}_{P})|X+\frac{1}{2}x_{\hbar}t,\pi_{P}+x_{\hbar}/\hbar\rangle.\\
 & =x_{\hbar}|X+\frac{1}{2}x_{\hbar}t,\pi_{P}+x_{\hbar}/\hbar\rangle
\end{align}

As described above, the quantum measurement also affects the $\hat{\pi}_{P}$
side.

\section*{Appendix C. The Kraus operator}

We discuss the Kraus operator in classical mechanics, which is used
in more modern measurement theory\footnote{Notation follows Section 5.}.
It has not been discussed in KvN formalism until now.

The Kraus operator is obtained by integrating out the measurement
device.

In quantum mechanics, a state $|\psi(t)\rangle$ is given by

\begin{equation}
|\psi(t)\rangle=\int dX|X\rangle\langle X|\hat{U}(t)|\phi\rangle|\eta\rangle=\int dX\hat{M}(X,t)|\phi\rangle|X\rangle,
\end{equation}
where

\begin{equation}
\hat{M}(X,t)\equiv\langle X|\hat{U}(t)|\eta\rangle=\int dx\langle X-x|\eta\rangle|x\rangle\langle x|.
\end{equation}

Positive Operator Valued Measure (POVM) $\hat{E}$ is constructed
as

\begin{equation}
\hat{E}(X,t)=\hat{M}^{\dagger}(X,t)\hat{M}(X,t).
\end{equation}

The probability of projective measurement of the measurement device
is
\begin{equation}
\mathrm{Pr}(X)=\langle\psi(t)|X\rangle\langle X|\psi(t)\rangle=\langle\psi(t)|\hat{E}(X,t)|\psi(t)\rangle.
\end{equation}

The state after measurement is
\begin{equation}
\hat{M}(X,t)|\phi\rangle|X\rangle.
\end{equation}

On the other hand, integrating out with $P$ gives another $\hat{M}$,
\begin{equation}
\hat{M}(P,t)\equiv\langle P|\hat{U}|\eta\rangle=e^{-i\hat{x}P}\langle P|\eta\rangle
\end{equation}

\inputencoding{latin9}%
In classical mechanics, a state $|\psi(t)\rangle$ is given by
\begin{equation}
|\psi(t)\rangle=\int dXdP|X,P\rangle\langle X|\hat{U}(t)|\phi\rangle|\eta\rangle=\int dXdP\hat{M}(X,P,t)|\phi\rangle|X,P\rangle,
\end{equation}
where

\begin{align}
\hat{M}(X,P,t) & \equiv\langle X,P|\hat{U}(t)|\eta\rangle=\int dxdp\langle X-x,P|\eta\rangle|x,p\rangle\langle x,p-P|.
\end{align}

In the same way,
\begin{equation}
\hat{M}(X, \pi_{P},t)=\int dxd\pi_{p}\langle X+x,\pi_{P}-\pi_{p}|\eta\rangle|x,\pi_{p}\rangle\langle x,\pi_{p}|,
\end{equation}

\begin{equation}
\hat{M}(\pi_{X}, P,t)=e^{iP\hat{\pi}_{p}-i\hat{x}\pi_{X}}\langle\pi_{X},P|\eta\rangle,
\end{equation}

\begin{equation}
\hat{M}(\pi_{X},\pi_{P},t)=\int dxd\pi_{p}\langle\pi_{X},\pi_{P}-\pi_{p}|\eta\rangle|x,\pi_{p}\rangle\langle x+\pi_{X}.\pi_{p}|.
\end{equation}

From the form of these expressions, $\hat{M}(X,t)$ in quantum mechanics
corresponds to $\hat{M}(X,\pi_{P},t)$ in classical mechanics and
$\hat{M}(P,t)$ in quantum mechanics corresponds to $\hat{M}(\pi_{X},P,t)$.

There is no counterpart to $\hat{M}(X,P,t),\hat{M}(\pi_{X},\pi_{P},t)$
in quantum mechanics.

\section*{Appendix D. The case of probability}

Consider the case where the initial condition is the only known probability.

\begin{equation}
|\psi(0)\rangle=|\phi\rangle|\eta\rangle,
\end{equation}

\begin{align}
\langle \phi,\eta|\hat{N}|\phi,\eta\rangle=\int dx\,dp\,dX\,dP(X+(t-1)x)|\phi(x,p)|^{2}|\eta(X,P)|^{2} \\
=\langle X\rangle{\eta}+(t-1)\langle x\rangle_{\phi},
\end{align}

\begin{equation}
\langle\phi,\eta|\hat{P}|\phi,\eta\rangle=\int dx\,dp\,dX\,dP tP|\phi(x,p)|^{2}|\eta(X,P)|^{2}=t\langle P\rangle_{\eta},
\end{equation}

Then, the unbiased condition is
\begin{equation}
\langle X\rangle_{\eta}=(1-t)\langle x\rangle_{\phi},\langle P\rangle_{\eta}=0.
\end{equation}

Under this condition,
\begin{equation}
\epsilon=\eta=0,
\end{equation}

\begin{equation}
\sigma(p)=\sigma(x)=0.
\end{equation}

If the unbiased condition is not satisfied, we get at $t=1,$

\begin{equation}
\epsilon=\langle X\rangle_{\eta},\eta=\langle P\rangle_{\eta},
\end{equation}

\begin{equation}
\langle X\rangle_{\eta}\sigma(p)+\langle P\rangle_{\eta}\sigma(x)=0.
\end{equation}

\section*{Appendix E. Planck Operator}
In quantum mechanics, $\hat{x}_{q}$ and $\hat{p}_{q}$ are commutative
rerlation,

\begin{equation}
[\hat{x}_{q},\hat{p}_{q}]=i\hbar.
\end{equation}
And time development of state is 
\begin{equation}
i\hbar\frac{d}{dt}|\psi\rangle=\hat{H}|\psi\rangle,
\end{equation}

where $\hat{H}$ is Hamiltonian.
Usualy, the Planck constant $\hbar$ is constants.
We suggest new quantum algebra
\begin{equation}
[\hat{x}_{q},\hat{p}_{q}]=i\hat{h},
\end{equation}

where $\hat{h}$ is operator. We call $\hat{\hbar}$ as ``Planck Operator''.

We introduce $\hat{\hbar}$'s conjugate operator $\hat{I}$, then,

\begin{equation}
[\hat{\hbar},\hat{I}]=i.
\end{equation}

We consider Planck constant as an operator because it introduces naturally
the relation between quantum mechanics and classical mechanics.

We introduce classical mechanics operators $\hat{x},\hat{p},\hat{\pi}_{x},\hat{\pi}_{p}$
,  and commutative relations,

\begin{equation}
[\hat{x},\hat{p}]=0,
\end{equation}

\begin{equation}
[\hat{x},\hat{\pi}_{x}]=i,
\end{equation}

\begin{equation}
[\hat{p},\hat{\pi}_{p}]=i.
\end{equation}

These constitute the KvN formalism of classical mechanics.

In classical mechanics, time development of state is
\begin{equation}
i\frac{d}{dt}|\psi_{c}\rangle=\hat{L}|\psi_{c}\rangle,
\end{equation}

where $\hat{L}$ is Liouvillian.

Now, we introduce commutative relations of $\hat{I}$ and $\hat{x}_{q}$
such as,

\begin{equation}
[\hat{I},\hat{x}_{q}]=\hat{\pi}_{p},
\end{equation}

\begin{equation}
[\hat{I},\hat{p}_{q}]=-\hat{\pi}_{x},
\end{equation}

\begin{equation}
[\hat{I},\hat{x}]=[\hat{I},\hat{p}]=[\hat{I},\hat{\pi}_{x}]=[\hat{I},\hat{\pi}_{p}]=0.
\end{equation}

We obtain
\begin{equation}
\hat{x}_{q}=\hat{x}+\hat{\hbar}\hat{\pi}_{p}
\end{equation}

\begin{equation}
\hat{p}_{q}=\hat{p}+\hat{\hbar}\hat{\pi}_{x}
\end{equation}

\begin{equation}
\hat{x}_{q}=e^{-\hat{\hbar}\hat{\pi}_{p}\hat{\pi}_{x}}\hat{x}e^{+\hat{\hbar}\hat{\pi}_{p}\hat{\pi}_{x}}
\end{equation}
\begin{equation}
\hat{p}_{q}=e^{-\hat{\hbar}\hat{\pi}_{p}\hat{\pi}_{x}}\hat{p}e^{+\hat{\hbar}\hat{\pi}_{p}\hat{\pi}_{x}}
\end{equation}

These relation introduce a natural relation of $\hat{L}$ and $\hat{H}$,
such as
\begin{equation}
\hat{L}=e^{\hat{\hbar}\hat{\pi}_{p}\hat{\pi}_{x}}[\hat{I},\hat{H}]e^{-\hat{\hbar}\hat{\pi}_{p}\hat{\pi}_{x}},
\end{equation}

because
\begin{equation}
e^{\hat{\hbar}\hat{\pi}_{p}\hat{\pi}_{x}}[\hat{I},\hat{H}(\hat{x}_{q},\hat{p}_{q})]e^{-\hat{\hbar}\hat{\pi}_{p}\hat{\pi}_{x}}=e^{\hat{\hbar}\hat{\pi}_{p}\hat{\pi}_{x}}\left(\frac{\partial\hat{H}}{\partial x_{q}}[\hat{I},\hat{x}_{q}]+\frac{\partial\hat{H}}{\partial p_{q}}[\hat{I},\hat{p}_{q}]\right)e^{-\hat{\hbar}\hat{\pi}_{p}\hat{\pi}_{x}}
\end{equation}

\begin{equation}
=e^{\hat{\hbar}\hat{\pi}_{p}\hat{\pi}_{x}}\left(\frac{\partial\hat{H}}{\partial x_{q}}\hat{\pi}_{p}-\frac{\partial\hat{H}}{\partial p_{q}}\hat{\pi}_{x}\right)e^{-\hat{\hbar}\hat{\pi}_{p}\hat{\pi}_{x}}
\end{equation}

\begin{equation}
=\frac{\partial\hat{H}(\hat{x},\hat{p})}{\partial x}\hat{\pi}_{p}-\frac{\partial\hat{H}(\hat{x},\hat{p})}{\partial p}\hat{\pi}_{x}=\hat{L}.
\end{equation}

We may also introduce thermal fluctattion. we may describe thermal
fluctuation and quantum fluctuation at once.

\section*{Appendix F. Time evolution of the von Neumann model}
Here,  we discuss the time dependence of the von Neumann model explicitly following\cite{key-Mello}.
\begin{equation}
V = \epsilon g(t)xP
\end{equation}

And If the interaction occurs only at $t=t_{1}$, 
$g(t)$ can be approximated as follows,

\begin{equation}
g(t)=\delta(t-t_{1}).
\end{equation}

 Therefore in quantum theory, using
\begin{equation}
\hat{V}(t)=\epsilon g(t)\hat{x}\hat{P},
\end{equation}
We obtain the Hamiltonian as
\begin{equation}
	\hat{H}=\hat{H}+\hat{V}(t),
\end{equation}
and the time evolution as
\begin{equation}
	i\hbar\frac{\partial}{\partial t}|\psi(t)\rangle=\hat{H}|\psi(t)\rangle.
\end{equation}

We set the initial state to
\begin{equation}
	|\psi(0)\rangle=|\phi\rangle|\eta\rangle.
\end{equation}
 
Therefore we describe the time evolution of the state as 

\begin{equation}
	|\psi\rangle_{I}=\hat{U}_{0}|\psi(t)\rangle\equiv\hat{U}_{I}(t)|\psi(0)\rangle,
\end{equation}

\begin{equation}
	\hat{U}_{I}(t)=\hat{U}_{0}^{\dagger}(t)\hat{U}(t),
\end{equation}

\begin{equation}
	i\hbar\frac{\partial}{\partial t}\hat{U}_{I}(t)=\hat{V}_{I}(t)\hat{U}_{I}(t),
\end{equation}

\begin{equation}
	\hat{V}_{I}(t)=\hat{U}_{0}^{\dagger}(t)\hat{V}(t)\hat{U}_{0}(t),
\end{equation}

\begin{equation}
	\hat{V}_{I}(t)=\epsilon\delta(t-t_{1})\hat{x}(t)\hat{P}(t)
\end{equation}

in the interaction picture.

At $t>t_{1}$, $\hat{U}(t)$ is

\begin{equation}
	\hat{U}_{I}(t)=e^{-\frac{i}{\hbar}\int_{0}^{t}\hat{V}_{I}(t')dt'}=e^{-\frac{i}{\hbar}\epsilon\hat{x}(t_{1})\hat{P}(t_{1})}=\hat{U}_{0}^{\dagger}(t_{1})e^{-\frac{i}{\hbar}\epsilon\hat{x}\hat{P}}\hat{U}_{0}(t_{1})=\hat{U}_{I,f}.
\end{equation}

Thus, if we convert this to a Schrodinger picture, we obtain
\begin{equation}
	\hat{U}_{f}=\hat{U}_{0}(t)\hat{U}_{I,f}=\hat{U}_{0}(t-t_{1})e^{-\frac{i}{\hbar}\epsilon\hat{x}\hat{P}}\hat{U}_{0}(t_{1}).
\end{equation}

We can understand $\hat{U}_f$ that the free motion until $t=t_{1}$, then the interaction $t=t_1$  and finally the free motion is performed.

This discussion can be repeated in classical mechanics of KvN formalism.

In this case, the Liouvillian $\hat{L}$ is
\begin{equation}
	\hat{L}=\hat{L}_{0}+\hat{L}_{1},
\end{equation}

\begin{equation}
	\hat{L}_{0}=\frac{\partial H_{0}}{\partial\hat{p}}\hat{\pi}_{x}-\frac{\partial H_{0}}{\partial\hat{x}}\hat{\pi}_{p}+\frac{\partial H_{0}}{\partial\hat{P}}\hat{\pi}_{X}-\frac{\partial H_{0}}{\partial\hat{X}}\hat{\pi}_{P},
\end{equation}

\begin{equation}
	\hat{L}_{1}=\frac{\partial V}{\partial\hat{p}}\hat{\pi}_{x}-\frac{\partial V}{\partial\hat{x}}\hat{\pi}_{p}+\frac{\partial V}{\partial\hat{P}}\hat{\pi}_{X}-\frac{\partial V}{\partial\hat{X}}\hat{\pi}_{P}=\epsilon\delta(t-t_{1})\left(\hat{P}\hat{\pi}_{p}-\hat{\pi}_{X}\hat{x}\right).
\end{equation}

By the same argument as quantum mechanics, we obtain
\begin{equation}
	\hat{U}_{f}=\hat{U}_{0}(t)\hat{U}_{I,f}=\hat{U}_{0}(t-t_{1})e^{-\frac{i}{\hbar}\epsilon\left(\hat{P}\hat{\pi}_{p}-\hat{\pi}_{X}\hat{x}\right)}\hat{U}_{0}(t_{1}).
\end{equation}

It shows that, like quantum mechanics, we can regard $U_f$ as performing free motion until $t=t_{1}$, interacting at $t=t_{1}$, and after that performing the free motion.

\section*{Acknowledgments}

The author thanks Takahiro Tsuchida for reading the paper and inviting
him to the meeting. He is deeply grateful to Tsukasa Yumibayashi for
providing early feedback. He is indebted to Akio Sugamoto and Shiro
Komata for reading this paper and giving useful comments. He sincerely
thanks Izumi Tsutsui, Lee Jaeha, and Yuichiro Mori, Yuki Inoue for
useful comments. He thanks Fumio Hiroshima for pointing out the von
Neumann's uniqueness theorem.

\end{document}